\documentclass[pre,12pt,notitlepage,tightenlines,superscriptaddress]{revtex4}
\pdfoutput=1

\usepackage{amsmath}
\usepackage{amsfonts}
\usepackage{bm}
\usepackage{euscript}
\usepackage{graphicx}
\usepackage{wasysym}
\usepackage{color}
\usepackage{subfigure}

\graphicspath{{figs/}}

\newcommand{\mathnotation}[2]{\newcommand{#1}{\ensuremath{#2}}}

\renewcommand{\l}{\left}               
\renewcommand{\r}{\right}              
\mathnotation{\pd}{\partial}           
\mathnotation{\ee}{{\mathrm e}}        
\mathnotation{\imi}{\mathrm{i}}        
\mathnotation{\ldef}{\mathrel{\raisebox{.069ex}{:}\!\!=}}
\mathnotation{\rdef}{\mathrel{=\!\!\raisebox{.069ex}{:}}}
\mathnotation{\dint}{\,{\mathrm{d}}}   
\mathnotation{\dVol}{V}                
\mathnotation{\darea}{a}               
\mathnotation{\cc}{\mathrm{c.c.}}

\mathnotation{\grad}{\nabla}           
\mathnotation{\lapl}{\Delta}           
\renewcommand{\div}{\grad\cdot}        
\mathnotation{\sdim}{d}                

\renewcommand{\time}{t}                
\mathnotation{\pdt}{\partial_\time}    
\mathnotation{\rv}{\bm{\xc}}           
\mathnotation{\xc}{x}                  
\mathnotation{\yc}{y}                  
\mathnotation{\zc}{z}                  
\mathnotation{\ixc}{\xi}               
\mathnotation{\uv}{\bm{\uc}}           
\mathnotation{\uc}{u}                  
\mathnotation{\vc}{v}                  
\mathnotation{\wc}{w}                  
\mathnotation{\Lx}{h}                  
\mathnotation{\Ly}{L_\yc}              
\mathnotation{\Lz}{L_\zc}              

\mathnotation{\xuv}{\bm{\hat\xc}}
\mathnotation{\yuv}{\bm{\hat\yc}}

\mathnotation{\Us}{U}                  
\mathnotation{\Ls}{L}                  
\mathnotation{\ci}{c_1}                
\mathnotation{\cii}{c_2}               
\mathnotation{\ciii}{c_3}              
\mathnotation{\civ}{c_4}               
\mathnotation{\phix}{\Phi}             

\mathnotation{\Fv}{\bm{F}}             
\mathnotation{\Fx}{\mathcal{F}}        
\mathnotation{\Hflux}{H}               
\mathnotation{\Diff}{\kappa}           
\mathnotation{\lavg}{\langle}          
\mathnotation{\ravg}{\rangle}          
\mathnotation{\Pe}{\mathrm{Pe}}        
\mathnotation{\MR}{\Gamma}             
\mathnotation{\un}{\hat{\bm{n}}}       
\mathnotation{\LLt}{\mathcal{L}^*}     

\mathnotation{\fq}{\theta}             
\mathnotation{\F}{\Theta}              
\mathnotation{\Xc}{X}                  
\mathnotation{\Yc}{Y}                  
\mathnotation{\qc}{q}                  
\mathnotation{\Qc}{Q}                  
\mathnotation{\Rrt}{R^2}               
\mathnotation{\vdecay}{\nu}            
\mathnotation{\pang}{\gamma}           
\mathnotation{\Uc}{\uc}

\mathnotation{\Sw}{S}                  
\mathnotation{\Swr}{S_{\mathrm{r}}}    
\mathnotation{\Swi}{S_{\mathrm{i}}}    
\mathnotation{\Lmix}{L_{\mathrm{mix}}} 
\mathnotation{\ucinf}{\uc}             
\mathnotation{\Hfluxinf}{\Hflux_{\infty}}
\mathnotation{\Peinf}{\Pe_{\infty}}    

\newcommand{\atx}[2]{{\l.#1\r\rvert}_{#2}}
\newcommand{\csa}[1]{\overline{#1}}
\newcommand{\csaev}[2]{\atx{\csa{#1}}{#2}}

\newcommand{\wiscaffil}{\affiliation{Department of Mathematics,
    University of Wisconsin, Madison, WI, USA}}
\newcommand{\michaffil}{\affiliation{Departments of Mathematics,
    Physics, and Center for Study of Complex Systems, University of
    Michigan, Ann Arbor, MI, USA}}
\newcommand{\imaaffil}{\affiliation{Institute for Mathematics and
    Applications, University of Minnesota, Minneapolis, MN, USA}}

\newcommand{\Order}[1]{\mathrm{O}\!\l(#1\r)}
\newcommand{\intV}[1]{\int_\MR #1 \dint\dVol} 

\begin{document}

\title{The Mixing Efficiency of Open Flows}

\wiscaffil
\michaffil
\imaaffil

\author{Jean-Luc Thiffeault}
\email{jeanluc@math.wisc.edu}
\wiscaffil
\imaaffil
\author{Charles R. Doering}
\email{doering@umich.edu}
\michaffil
\imaaffil



\date{\today}

\begin{abstract}
  Mixing in open incompressible flows is studied in a model problem
  with inhomogeneous passive scalar injection on an inlet boundary.
  As a measure of the efficiency of stirring, the bulk scalar
  concentration variance is bounded and the bound is shown to be sharp
  at low P\'eclet number.  Although no specific flow saturating the
  bound at high P\'eclet number is produced here, the estimate is
  conjectured to be approached for flows possessing sufficiently
  sustained chaotic regions.
\end{abstract}

\maketitle

\section{Introduction}

Since the pioneering work of L.\ N.\ Howard~\cite{Howard1963}, applied
physicists and mathematicians have found that rigorous bounds are a
straightforward way of obtaining useful estimates on fluxes of
quantities transported by complex incompressible flows, from laminar
to fully developed turbulent flows.  In recent years the co-authors and
collaborators~\cite{Thiffeault2004, DoeringThiffeault2006, Shaw2007,
  Thiffeault2008, Okabe2008, ONaraigh2008} and
others~\cite{Balkovsky1999, Plasting2006, Birch2007, Keating2009_preprint,
  Turner2009} used similar ideas to bound mixing efficiencies for
processes described by the advection-diffusion equation with
inhomogeneous passive scalar sources and sinks.  One major insight
emerging from these studies is the distinction between transient
mixing problems, where concentration inhomogeneities are provided by
initial conditions, and (statistically) steady mixing in the presence
of sustained sources and sinks.  In the former case the flow's
shearing and straining properties, i.e., the stretching and folding of
material lines, serve to enhance the effect of molecular diffusion and
increase the rate of scalar concentration variance dissipation.  On
the other hand, in the latter case the spatial variation of the
concentration is best suppressed by direct transport of material from
sources to sinks and vice versa.  Then the maximum possible
enhancement of molecular diffusion by stirring is limited not by small
scale features in the flow, but rather by small scale properties of
the sources and sinks~\cite{DoeringThiffeault2006, Shaw2007,
  Okabe2008}.

In this paper we extend these techniques to the case of fluxes at the
boundaries rather than body sources as previously described.  This is
not just a mathematical detail: the problem changes significantly.
Notably, the scalar inhomogeneities are transported away from both
source and sink regions so the transient mixing efficiency of
downstream regions of the stirring flow are crucial for efficient
mixing.  At the same time the source-sink structure defined by the
distribution of fluxes on the boundary plays a key role in the maximum
possible mixing efficiency.  In the remainder of this paper we define
a specific scenario where these issues can be addressed and study it
via general analysis and particular examples.

\section{A Bound on the Variance}

The suppression of concentration variance is a reasonable measure of
the efficiency of mixing since in the absence of sources it decays due
to diffusive mixing and it vanishes when complete homogenization is
achieved.  The variance is also a useful measure in the context of
open flows~\cite{Danckwerts1953, Gouillart2009,
  Gouillart2010_preprint}.  We use techniques developed
in~\cite{Thiffeault2004,DoeringThiffeault2006,Shaw2007,Okabe2008,ONaraigh2008}
to bound the time-averaged variance of the concentration of a passive
scalar.  We shall use both the variance and its square root, the
standard deviation, but final results will usually be quoted in terms
of the variance.

\subsection{The model and definitions}

Consider a fixed domain~$\MR$, in which flows a fluid with
\begin{figure}
  \includegraphics[width=.6\textwidth]{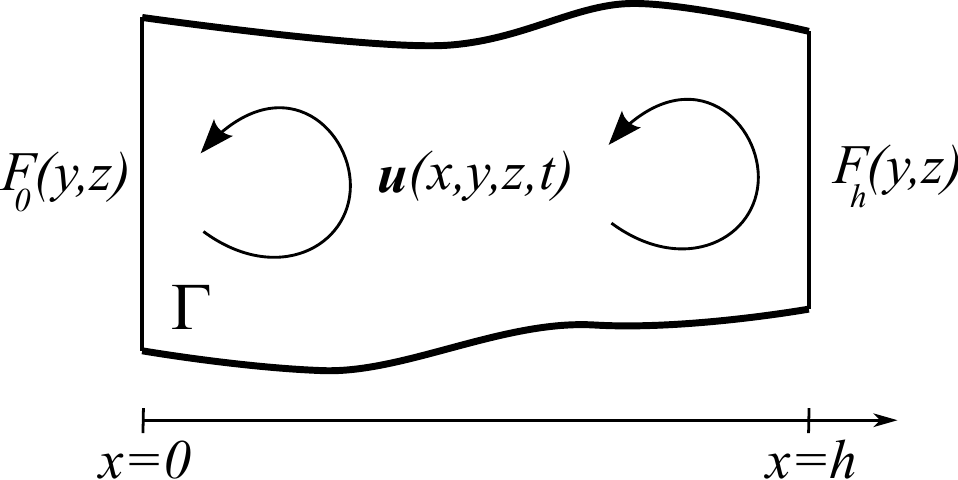}
  \caption{Schematic of the flow domain~$\MR$.  The scalar flux is
    imposed at~$\xc=0$ and~$\xc=\Lx$, and there is no flux on the
    sidewalls.}
  \label{fig:domain}
\end{figure}
incompressible velocity field~$\uv(\rv,\time)$.  The fluid transports
a diffusing passive scalar with concentration~$\theta(\rv,\time)$
which is stirred to some degree of homogeneity within~$\MR$.  The
concentration~$\theta(\rv,\time)$ in the domain~$\MR$ obeys the
advection-diffusion equation
\begin{equation}
  \pdt\theta + \uv\cdot\grad\theta = \Diff\lapl\theta.
  \label{eq:AD}
\end{equation}
The flux of~$\theta(\rv,\time)$ is defined by
\begin{equation}
  \Fv \ldef \uv\,\theta - \Diff\grad\theta\,.
  \label{eq:Fvdef}
\end{equation}
At the boundary~$\pd\MR$ of the domain, we specify
the normal flux $\Fv\cdot\un$ as
\begin{equation}
  (\Fv\cdot\un)(0,\yc,\zc) = \Fx_0(\yc,\zc),\quad
  (\Fv\cdot\un)(\Lx,\yc,\zc) = \Fx_\Lx(\yc,\zc),\quad
  (\Fv\cdot\un)_{\text{sidewalls}}=0,
  \label{eq:BC}
\end{equation}
where the sidewalls are the curved sections in Fig.~\ref{fig:domain}.
We denote the horizontal flux~$\Fx(\rv)\ldef \xuv\cdot\Fv(\rv)$, and
we will often write~$\Fx_\xc(\yc,\zc)$ to mean $\Fx(\xc,\yc,\zc)$, as
in the boundary condition~\eqref{eq:BC} above.  Note that there is
considerable literature, and a long running debate, regarding the
appropriate boundary conditions to use for open-flow
systems~\cite{Hulburt1944, Danckwerts1953, Pearson1959, Nauman1981,
  Hisaka1999}; we use a prescribed normal flux for mathematical
expediency.

Multiply the advection-diffusion equation~\eqref{eq:AD} by an
arbitrary smooth test function~$\varphi(\rv,\time)$ and integrate:
\begin{equation}
  \intV{\varphi\,\pdt\theta} + \intV{\varphi\div(\uv\,\theta)} =
  \Diff\intV{\varphi\lapl\theta}\,
  \label{eq:ADavg}
\end{equation}
where~$\dint\dVol=\dint\xc\dint\yc$ in 2D
and~$\dint\xc\dint\yc\dint\zc$ in 3D.  To obtain~\eqref{eq:ADavg}, we
have also used the incompressibility condition~$\div\uv=0$.  Next
follow a few integration by parts,
\begin{multline}
  \pdt\intV{\theta\,\varphi} - \intV{\theta\pdt\varphi} +
  \intV{\div(\uv\,\varphi\,\theta)} -
  \intV{\theta\,\div(\uv\,\varphi)} \\ =
  \Diff\l(\intV{\div(\varphi\grad\theta)}\,
  - \intV{\div(\theta\,\grad\varphi)}\,
  + \intV{\theta\lapl\varphi}\r).
  \label{eq:avg2}
\end{multline}
We collect on the right the boundary terms,
\begin{equation}
  \pdt\intV{\theta\,\varphi} - \intV{\theta\l(\pdt\varphi
  + \div(\uv\,\varphi)
  + \Diff\lapl\varphi\r)}\,
  =
  -\intV{\div(\varphi\,\Fv + \Diff\,\theta\,\grad\varphi)}\,,
  \label{eq:avg3}
\end{equation}
and turn them into a surface integral,
\begin{equation}
  \text{RHS} = -
  \int_{\pd\MR}(\varphi\,\Fv
  + \Diff\,\theta\,\grad\varphi)\cdot\un\dint\darea\,,
  \label{eq:surfint}
\end{equation}
where~$\dint\darea=\dint\yc$ in 2D and~$\dint\yc\dint\zc$ in 3D.  On
the sidewalls, we have~$\Fv\cdot\un=0$ because of boundary
conditions~\eqref{eq:BC}, so the flux term in~\eqref{eq:surfint}
vanishes.  Since we haven't solved the full problem~(\ref{eq:AD}), we
don't know~$\theta$ on the boundary.  To remove the direct~$\theta$
dependence in~\eqref{eq:surfint}, we restrict attention to test
functions~$\varphi$ with
\begin{equation}
  \grad\varphi\cdot\un=0 \quad \text{on} \quad \pd\MR.
\end{equation}
All that remains of the integral~\eqref{eq:surfint} are the surfaces
at~$\xc=0$ and~$\xc=\Lx$, and Eq.~\eqref{eq:avg3} is then
\begin{equation}
  \pdt\intV{\theta\,\varphi} + \lavg\theta\,\LLt\varphi\ravg
  =
  \int_{\pd\MR_0}{\varphi\,\Fx}_{0}\dint\darea
  - \int_{\pd\MR_\Lx}{\varphi\,\Fx}_{\Lx}\dint\darea,
  \label{eq:avg4}
\end{equation}
with~$\dint\darea=\dint\yc$ in 2D and~$\dint\yc\dint\zc$ in 3D, and
$\pd\MR_\xc$ is the cross-section at constant~$\xc$.  We also defined
the formal adjoint of the advection-diffusion operator,
\begin{equation}
  \LLt\varphi \ldef
  - \pdt\varphi - \div(\uv\,\varphi) - \Diff\lapl\varphi\,.
  \label{eq:Ladj}
\end{equation}

Equation~\eqref{eq:avg4} is the fundamental `integrated equation' that
we will be using subsequently.  Setting~$\varphi=1$
in~\eqref{eq:avg4}, we get an equation for the conservation of
total~$\theta$,
\begin{equation}
  \pdt\intV{\theta} =
  \int_{\pd\MR_0}{\Fx}_{0}\dint\darea
  - \int_{\pd\MR_\Lx}{\Fx}_{\Lx}\dint\darea
  \label{eq:avgtheta}
\end{equation}
which of course says that the change in~$\intV{\theta}$ is
the net difference in fluxes at~$\xc=0$ and~$\xc=\Lx$.  For the
remainder of the paper, we shall consider the case where
\begin{equation}
  \int_{\pd\MR_0}{\Fx}_{0}\dint\darea
  = \int_{\pd\MR_\Lx}{\Fx}_{\Lx}\dint\darea = 0
\end{equation}
so that the total amount of~$\theta$ in~$\MR$ is always~$0$.

\subsection{Lower bound on the variance}

We define angle brackets and overbars as
\begin{equation}
  \lavg f \ravg \ldef
  \lim_{T \rightarrow \infty} \frac{1}{T} \int_0^T\!\!\dint\time
  \int_\MR f(\rv,\time)\, \dint\dVol,\qquad
  \csa{f} \ldef
  \lim_{T \rightarrow \infty} \frac{1}{T} \int_0^T\!\!\dint\time
  \int_{\pd\MR_\xc}f(\rv,\time)\dint\darea.
  \label{eq:intdef2}
\end{equation}
(The limits are assumed to converge.)  After time-integrating
Eq.~\eqref{eq:avg4} the term~$\lavg\pdt(\varphi\,\theta)\ravg$
vanishes and we have
\begin{equation}
  \lavg\theta\,\LLt\varphi\ravg
  =
  {\csaev{(\varphi\,\Fx
      + \Diff\,\theta\,\pd_\xc\varphi)}{0}}
  - {\csaev{(\varphi\,\Fx
      + \Diff\,\theta\,\pd_\xc\varphi)}{\Lx}}\,.
  \label{eq:tavg}
\end{equation}
The notation~$\csaev{f}{c}$ signifies~$\csa{f}$ evaluated at~$\xc=c$.
The conservation equation~\eqref{eq:avgtheta}
implies~\hbox{$\lavg\theta\ravg=0$}, so the spatially-integrated
variance of the concentration field is~$\lavg\theta^2\ravg$.  We shall
omit the qualifier `spatially-integrated' most of the time.

Using the Cauchy--Schwartz inequality, we can extract from~\eqref{eq:tavg} a
bound on the standard deviation,
\begin{equation}
  \lavg\theta^2\ravg^{1/2}
  \ge
  \frac{\l\lvert\l(\csaev{\varphi\,\Fx}{0}
    - \csaev{\varphi\,\Fx}{\Lx}\r)\r\rvert}
  {\lavg\lvert\LLt\varphi\rvert^2\ravg^{1/2}}\,.
  \label{eq:thebound}
\end{equation}
The right-hand side of the bound~\eqref{eq:thebound} contains only
known quantities.  The bound says that for given functions~$\uv$,
$\Fx_0\ldef\atx{\Fx}{0}$,~$\Fx_\Lx\ldef\atx{\Fx}{\Lx}$, and
diffusivity~$\Diff$, the passive scalar must exhibit a minimum level of
fluctuations.  Since better mixing requires decreasing fluctuations,
Eq.~\eqref{eq:thebound} is a rigorous bound on the potential efficiency of
a mixer.  In practice, one tries to find a function~$\varphi$ to make
the bound as sharp as possible, that is to maximize the right-hand
side of~\eqref{eq:thebound}.  The optimal $\varphi$ will, in
principle, depend on all the problem parameters.

Observe that adding a constant~$c$ to~$\varphi$ changes the numerator
in~\eqref{eq:thebound} by a factor~$c\,(\csa{\Fx_0} - \csa{\Fx_\Lx})$,
which vanishes because the total amount of~$\theta$ within the domain
is conserved (after the long-time averaging).  Hence, adding a
constant to~$\varphi$ does not affect the estimate.

If we have a semi-infinite `channel,' with~$\Lx\rightarrow\infty$,
then the scalar field at~$\Lx$ will become completely homogeneous due
to diffusion ($\theta\rightarrow 0$) and its flux will necessarily
vanish.  We then have
\begin{equation}
  \lavg\theta^2\ravg^{1/2}
  \ge
  \frac{\l\lvert(\csaev{\varphi\,\Fx}{0})\r\rvert}
  {\lavg\lvert\LLt\varphi\rvert^2\ravg^{1/2}}\,.
  \label{eq:infbound}
\end{equation}
Equation~\eqref{eq:infbound} is ideal for this channel configuration,
since we only require knowledge of the flux at~$\xc=0$.
(Alternatively, we could also obtain~\eqref{eq:infbound} by
specifying~$\varphi=0$ at~$\xc=\Lx$.)  We shall focus exclusively on
the semi-infinite channel configuration ($\Lx\rightarrow\infty$) for
the remainder of the paper.

\subsection{A bound for all flows}
\label{sec:globalbound}

The bound~\eqref{eq:infbound} requires complete knowledge of the
velocity field~$\uv$.  However, we can turn it into a `global' bound
over all allowable velocity fields.  For simplicity, from now on we
restrict attention to a uniform, rectangular cross section of
dimensions~$\Ly$ and~$\Lz$, and to a time-independent velocity field.

First we bound the numerator in~\eqref{eq:infbound}.  For a
time-independent test function~$\varphi$,
\begin{equation}
  \lavg\lvert\LLt\varphi\rvert^2\ravg
  = \lavg\lvert\uv\cdot\grad\varphi + \Diff\lapl\varphi\rvert^2\ravg
  \le 2\lavg\lvert\uv\cdot\grad\varphi\rvert^2\ravg
  + 2\Diff^2\lavg\lvert\lapl\varphi\rvert^2\ravg.
  \label{eq:Mink}
\end{equation}
We bound the first term in~\eqref{eq:Mink},
\begin{align*}
  2\lavg\lvert\uv\cdot\grad\varphi\rvert^2\ravg
  &\le 2\int_0^\infty\csa{\lvert\uv\rvert^2
    \lvert\grad\varphi\rvert^2}\dint\xc \\
  &\le 2\int_0^\infty\csa{\lvert\uv\rvert^2}
  \sup_{\yc,\zc}\,\lvert\grad\varphi\rvert^2\dint\xc \\
  &\le 2\sup_{\xc}\csa{\lvert\uv\rvert^2}\int_0^\infty
  \sup_{\yc,\zc}\,\lvert\grad\varphi\rvert^2\dint\xc
  = 2\Us^2\Ls^{\sdim-2}\,\ci
\end{align*}
where we defined the mixing flow velocity scale~$\Us\ge0$ by
\begin{equation}
  \Us^2 \ldef \Ls^{1-\sdim}\,\sup_{\xc}\csa{\lvert\uv\rvert^2}
  \label{eq:Usdef}
\end{equation}
and the length scale~$\Ls\ldef\sqrt{\Ly\Lz}$ in 3D, or $\Ls\ldef\Ly$
in 2D.  We also defined the dimensionless coefficient
\begin{equation}
  \ci\{\varphi\} \ldef \Ls
  \int_0^\infty \sup_{\yc,\zc}\,\lvert\grad\varphi\rvert^2\dint\xc\,.
\end{equation}
We define another dimensionless coefficient
\begin{equation}
  \cii\{\varphi\} \ldef \Ls^{4-\sdim}\,\lavg\lvert\lapl\varphi\rvert^2\ravg
\end{equation}
which, using~\eqref{eq:Mink}, allows us to write
\begin{equation}
  \lavg\lvert\LLt\varphi\rvert^2\ravg
  \le 2\Us^2\Ls^{\sdim-2}\,\ci + 2\Diff^2\Ls^{\sdim-4}\,\cii
  = 2\Diff^2\Ls^{\sdim-4}\l(\ci\,\Pe^2 + \cii\r)
\end{equation}
where the P\'eclet number is defined as
\begin{equation}
  \Pe \ldef \Us\Ls/\Diff.
  \label{eq:Pedef}
\end{equation}

Now we examine the numerator in~\eqref{eq:infbound}.  The flux~$\Fx$
has units of velocity (considering without loss of generality that
$\theta$ is dimensionless), so we can define the dimensionless
coefficient
\begin{equation}
  \ciii\{\varphi,\Fx_0\} \ldef
  \Us^{-1}\Ls^{1-\sdim}\,\l\lvert(\csaev{\varphi\,\Fx}{0})\r\rvert\,.
\end{equation}

Combining the results we can write the bound as
\begin{equation}
  \lavg\theta^2\ravg^{1/2}
  \ge
  \Ls^{\sdim/2}\frac{\ciii\,\Pe}
  {\sqrt{2(\ci\,\Pe^2 + \cii)}}\,,
  \label{eq:globalbound}
\end{equation}
where now the properties of the stirring flow only appear as the
scale~$\Us$ in~$\Pe$.  For small or large~$\Pe$, we can improve the
bound a little by neglecting the appropriate term in~\eqref{eq:Mink},
which allows us to leave out the factor of~$2$ in that equation.  We
then get the asymptotic estimates
\begin{equation}
  \lavg\theta^2\ravg^{1/2}
  \gtrsim
  \begin{cases}
    \Ls^{\sdim/2}\,({\ciii}/{\sqrt{\cii}})\,\Pe\,,\qquad & \Pe \ll 1; \\
    \Ls^{\sdim/2}\,({\ciii}/{\sqrt{\ci}}), & \Pe \gg 1.
  \end{cases}
  \label{eq:globalboundasym}
\end{equation}
For convenience of display, we combine the asymptotic
bounds~\eqref{eq:globalboundasym} into the same form
as~\eqref{eq:globalbound},
\begin{equation}
  \lavg\theta^2\ravg^{1/2}
  \gtrsim
  \Ls^{\sdim/2}\frac{\ciii\,\Pe}
  {\sqrt{\ci\,\Pe^2 + \cii}}\,,
  \quad \Pe\ll1 \text{ or } \Pe \gg1,
  \label{eq:globalboundapprox}
\end{equation}
where it is understood that this estimate is only valid asymptotically
for small or large~$\Pe$.  For our purposes, the error made at
intermediate~$\Pe$ will be small.

The small~$\Pe$ bound says that the standard deviation can become
small in that case: this is the diffusion-dominated limit.  For
large~$\Pe$, the lower bound saturates and becomes independent
of~$\Pe$: we shall examine this behavior more closely in
Section~\ref{sec:meander}, and see if it can be realized in practice.

\subsection{Choosing the test function $\varphi$}
\label{sec:phichoice}

To get usable numbers out of~\eqref{eq:globalboundapprox} we need to
choose~$\varphi$ for a given~$\Fx_0$.  We specialize to two dimensions
($\sdim=2$) and
\begin{equation}
  \Fx_0(\yc) = \Us \Hflux \sin \qc\yc\,,
  \label{eq:flux0sin}
\end{equation}
where the constant~$\Hflux$ is a dimensionless measure of the flux.
We consider periodic boundary conditions in~$\yc$, which for our
purposes is equivalent to the no-flux conditions on the sidewalls
in~\eqref{eq:BC}.  The wavenumber~$\qc=\Qc/\Ls$, where~$\Qc$ is an
integer multiple of~$2\pi$ and~$\Ls=\Ly$.

To get the sharpest bound, our goal is to maximize the right-hand side
of~\eqref{eq:globalboundapprox} as much as possible. Hence we need a
test function $\varphi$ with a large projection on~$\Fx_0$, to
maximize~$\ciii$.  The most expedient approach is to take a separable
form for~$\varphi$ with the~$\yc$ dependence mimicking~$\Fx_0$'s:
\begin{equation}
  \varphi = \phix(\xc) \, \sin \qc\yc, \qquad \phix(0)=1.
\end{equation}
We can then easily compute
\begin{equation}
  \ciii = \Us^{-1}\Ls^{-1}\, \phix(0) \, \Us \Hflux \int_0^\Ls
  \sin^2\qc\yc \dint\yc
  = \tfrac12 \Hflux.
\end{equation}
and
\begin{equation}
  \cii = \Ls^{2}\,\lavg\lvert\lapl\varphi\rvert^2\ravg
  = \Ls^{2}\,\lavg\lvert(\phix'' - \qc^2\phix)\sin\qc\yc\rvert^2\ravg
  = \tfrac12\Ls^3 \int_0^\infty\lvert\phix'' - \qc^2\phix\rvert^2\dint\xc\,.
  \label{eq:ciiphix}
\end{equation}
The constant~$\ci$ is more problematic, since we need to evaluate the
supremum inside the integral:
\begin{equation}
  \ci = \Ls
  \int_0^\infty \sup_\yc\,\lvert\grad\varphi\rvert^2\dint\xc
  = \Ls
  \int_0^\infty \sup_\yc\,\l((\phix')^2\sin^2\qc\yc
  + \qc^2\phix^2\cos^2\qc\yc\r)\dint\xc\,.
\end{equation}
To find the supremum, take the derivative and set to zero,
\begin{equation}
  \frac{d}{d\yc}\l((\phix')^2\sin^2\qc\yc
  + \qc^2\phix^2\cos^2\qc\yc\r)
  = \l((\phix')^2 - \qc^2\phix^2\r)\qc\sin 2\qc\yc = 0.
\end{equation}
We need only consider the extrema at~$\qc\yc = 0$ and~$\pi/2$,
since~$\sin 2\qc\yc$ function is~$\pi$-periodic, which leads to
\begin{equation}
  \ci
  = \Ls
  \int_0^\infty \max\l(\qc^2\phix^2\,,\,\phix'^2\r)\dint\xc\,.
  \label{eq:ciphix}
\end{equation}

Now we make specific choices for~$\phix$.  Consider first the limit of
small~$\Pe$.  In that case we ignore~$\ci$ and try to minimize~$\cii$.
The Euler--Lagrange equation for~\eqref{eq:ciiphix} is
\begin{equation}
  \phix'''' - 2\qc^2\phix'' + \qc^4\phix = 0.
\end{equation}
The appropriately integrable solution that also
satisfies~$\phix(0)=1$, $\phix'(0)=0$ is
\begin{equation}
  \phix(\xc) = (1 + \qc\xc)\,\ee^{-\qc\xc}
\end{equation}
which gives~$\cii = (\qc\Ls)^3 = \Qc^3$.

Conversely, for large $\Pe$ we ignore~$\cii$ and focus on~$\ci$.  In
that case, we might like to use the function~$\phix=\ee^{-\qc\xc}$,
for which $\phix'=-\qc\phix$, so the two choices for the max
in~\eqref{eq:ciphix} are the same, and~$\ci=\tfrac12(\qc\Ls) =
\tfrac12\Qc$ for the constant.  The only problem is that this choice
for~$\phix$ does not satisfy the boundary condition~$\phix'(0)=0$.
This is easily remedied by inserting a small boundary layer
near~$\xc=0$, and letting the size of the layer go to zero.  For
large~$\Pe$, the correction incurred can be neglected.

We can combine our two asymptotic results as
in~\eqref{eq:globalboundapprox} to conclude
\begin{equation}
  \frac{1}{\Ls \Hflux}\,\lavg\theta^2\ravg^{1/2}
  \gtrsim
  \frac{\Pe}{2\sqrt{\tfrac12\Qc\,\Pe^2 + \Qc^3}}\,,\qquad
  \quad \Pe\ll1 \text{ or } \Pe \gg1.
  \label{eq:globalboundnum}
\end{equation}
This estimate is rigorous for large and small~$\Pe$, and we
expect that error made at intermediate~$\Pe$ is typically small.  This
is illustrated by the example developed in the next section.

\section{Example: The Barberpole Flow Mixer}
\label{sec:barberpole}

As an illustration of typical features of open-flow systems, we study
a toy problem that can be solved exactly.  We will then compare the
exact standard deviation to the bound~\eqref{eq:globalboundnum}.  We
consider the 2D spatially uniform flow
\begin{equation}
  \uv = \Uc\,\xuv + \vc\,\yuv\,,
  \label{eq:bpflow}
\end{equation}
where~$\Uc$ and~$\vc$ are constants.  We we solve the steady
advection-diffusion equation for~$\theta(\rv)$, where~$\rv = (\xc,\yc)
\in \MR = [0,\infty)\times[0,\Ly]$, with periodic boundary conditions
in~$\yc$ and the flux~\eqref{eq:Fvdef} specified as
\begin{equation}
  \xuv\cdot\Fv(0,\yc) = \Fx_0(\yc),\qquad
  \Fv(\xc,\yc) \rightarrow 0 \quad\text{as}\quad \xc \rightarrow \infty.
  \label{eq:BCbp}
\end{equation}

For the case of pure advection, with~$\Diff=0$ in Eq.~\eqref{eq:AD},
the solution is
\begin{equation}
  \theta(\xc,\yc) =
    \Uc^{-1}\Fx_0\l(\yc - \xc\,\vc/\Uc\r)\,.
\end{equation}
The concentration has a `barberpole' pattern (see
Fig.~\ref{fig:bp_decay}): the imposed distribution at~$\xc=0$ is
twisted around the periodic direction~$\yc$, with the tilt
angle~$\pang$ determined by the ratio of~$\vc$ and~$\Uc$.  The
cross-sectional variance~$\csa{\theta^2}/\Ly$ is independent of~$\xc$,
and there is no mixing, as expected.

\subsection{Advection and diffusion}
\label{sec:barberpoleAD}

Now we include diffusion in Eq.~\eqref{eq:AD}.  We first
Fourier-expand in~$\yc$,
\begin{equation}
  \theta(\xc,\yc) = \sum_\qc \fq_\qc(\xc)\,\ee^{\imi\qc\yc} + \cc
  \label{eq:Fourierexp}
\end{equation}
where~$\qc$ is an integer multiple of~$2\pi/\Ly$.  In the steady case, the
advection-diffusion equation~\eqref{eq:AD} for the amplitude of the
vertical mode~$\qc$ becomes
\begin{equation}
  \Diff\fq_\qc'' - \Uc\,\fq_\qc' - \qc(\Diff\qc + \imi\,\vc)\fq_\qc = 0.
  \label{eq:ADsinglemode}
\end{equation}
Using the length scale~$\Ls=\Ly$ and velocity
scale~$\Us=\sqrt{\uc^2+\vc^2}$ (consistent with the definitions in
Section~\ref{sec:globalbound}), we define the rescaled
variables~$\Xc=\xc/\Ls$, $\Yc=\yc/\Ls$, $\Qc=\qc\Ls$,
and~$\F(\Xc)=\fq_\qc(\Ls\Xc)$ to get the dimensionless ODE
\begin{equation}
  \F'' - \Pe\,\cos\pang\,\F' - \Qc(\Qc + \imi\,\Pe\,\sin\pang)\F = 0
\end{equation}
where the pitch angle~$\pang$ is defined by~$\tan\pang=\vc/\uc$ and
the P\'eclet number~$\Pe$ is as in Eq.~\eqref{eq:Pedef}.  This has
solution
\begin{equation}
  \F(\Xc) = \F(0)\,\ee^{-\l(\vdecay + \imi\omega\r)\Xc},\qquad \Xc\ge 0,
  \label{eq:sol}
\end{equation}
where we defined the spatial decay rate by
\begin{equation}
  \vdecay = \alpha - \tfrac12\Pe\,\cos\pang
  \label{eq:vdecay}
\end{equation}
with
\begin{gather*}
  \alpha = \sqrt{\tfrac{1}{2}\l(\Rrt
    + \l(\tfrac{1}{4}\,\Pe^2\cos^2\pang + \Qc^2\r)\r)}\,,
  \qquad
  \omega = \sqrt{\tfrac{1}{2}\l(\Rrt
    - \l(\tfrac{1}{4}\,\Pe^2\cos^2\pang + \Qc^2\r)\r)},\\
  \Rrt = \sqrt{\Qc^2\,\Pe^2\sin^2\pang
    + \l(\tfrac{1}{4}\,\Pe^2\cos^2\pang + \Qc^2\r)^2}\,.
\end{gather*}
We have~$\nu>0$ since~$\alpha \ge \l(\tfrac{1}{4}\,\Pe^2\cos^2\pang +
\Qc^2\r)^{1/2} > \tfrac{1}{2}\,\Pe\,\cos\pang$; hence, the
solution~\eqref{eq:sol} decays with~$\Xc$, as required.  Using the
form~\eqref{eq:flux0sin} for the flux~$\Fx_0(\yc)$ and the boundary
condition~\eqref{eq:BCbp},
\begin{equation}
  \Fx_0(\yc) = \Us \Hflux \sin \qc\yc = \Fx(0,\yc)
  = \frac{\Diff}{\Ls}\l(\Pe\,\cos\pang
  + \l(\vdecay + \imi\omega\r)\r) \F(0)\,\ee^{\imi\qc\yc}
  + \cc
\end{equation}
which we can use to solve for
\begin{equation}
  \F(0) = \frac{1}{2\imi}\,\frac{\Hflux\,\Pe}
  {\Pe\,\cos\pang + \vdecay + \imi\omega}\,.
\end{equation}

\begin{figure}
  \includegraphics[width=.8\textwidth]{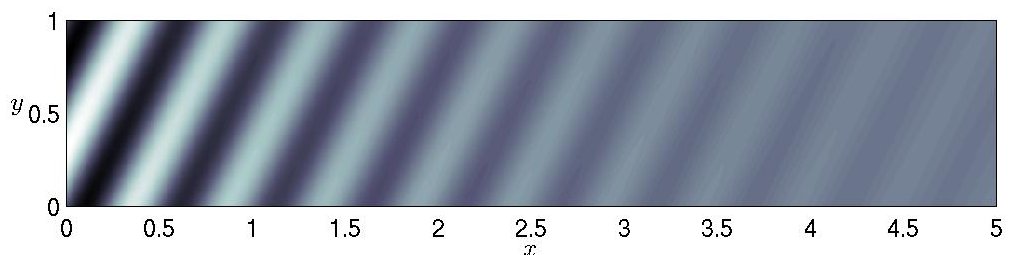}
  \caption{Concentration field~$\theta(\rv)$ for the barberpole flow
    (parameters values are~$\Hflux=1$, $\Pe = 100$, $\pang = 3\pi/8$,
    $\Ls=\Ly=1$).}
  \label{fig:bp_decay}
\end{figure}

Figure~\ref{fig:bp_decay} shows the concentration field
pattern for the parameter values~$\Hflux=1$, $\Qc=2\pi$, $\Pe = 100$,
$\pang = 3\pi/8$, $\Ls=\Ly=1$.  The tilted stripes make an
angle~$\pang$ with the horizontal, and the intensity decays in space
at a rate~$\vdecay$.

For small~$\Pe$, we have the expansions
\begin{equation}
  \vdecay = \Qc - \tfrac12\Pe\,\cos\pang + \Order{\Pe^2},\qquad
  \omega = \tfrac12\Pe\,\sin\pang + \Order{\Pe^3}.
\end{equation}
To leading order in~$\Pe$, the spatial decay rate~$\vdecay$ is
independent of~$\Pe$ and the pitch angle~$\pang$.  This is because in
the absence of flow (or for a weak flow) the decay scale length is set
by~$\qc$, and the diffusivity completely drops out.

\begin{figure}
  \includegraphics[width=.6\textwidth]{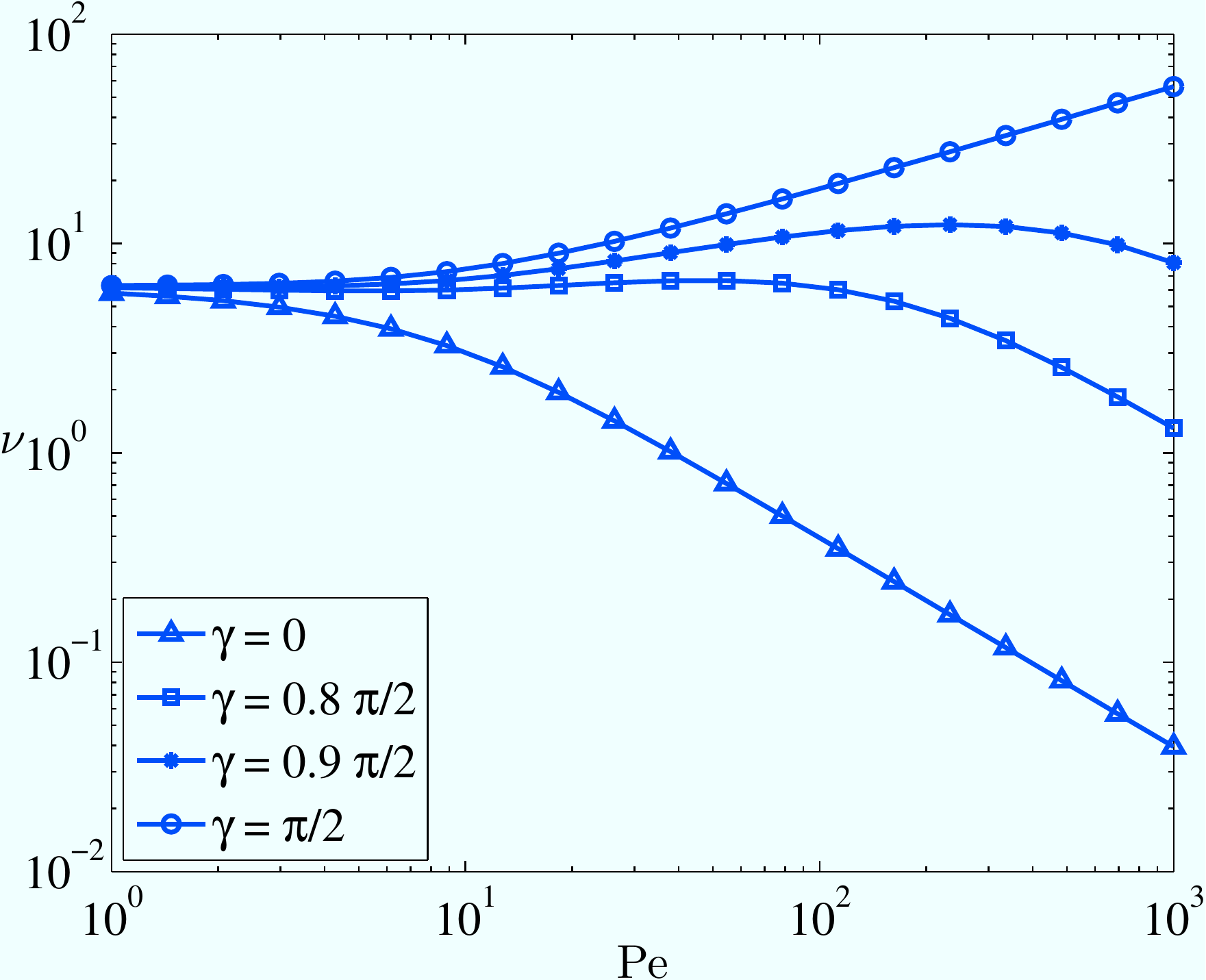}
  \caption{Decay rate~$\vdecay$ from Eq.~\eqref{eq:vdecay} for the
    barberpole flow, as a function of~$\Pe$.  For any value
    of~$\pang<\pi/2$, the decay rate ultimately goes to zero
    like~$\Pe^{-1}$ for~$\Pe\gg 1$.  However, for~$\pang=\pi/2$ the
    decay rate grows as~$\Pe^{1/2}$.}
  \label{fig:bp_vdecay}
\end{figure}

For large~$\Pe$, we have the expansions
\begin{equation}
  \vdecay = \Qc^2\sec^3\pang\,\Pe^{-1} + \Order{\Pe^{-3}},\qquad
  \omega = \Qc\tan\pang + \Order{\Pe^{-2}}.
\end{equation}
These are valid for~$\pang$ not too close to~$\pi/2$.
Figure~\ref{fig:bp_vdecay} shows the spatial decay rate~$\vdecay$ as
function of~$\Pe$, for several values of~$\pang$.  For large~$\Pe$,
the decay rate ultimately decays as~$\Pe^{-1}$.
However, for~$\pang=0$ the decay rate \emph{grows} as $\Pe^{1/2}$.
This case corresponds to no horizontal flow: the vertical flow is very
effective at pushing the source onto the sink, leading to fast spatial
decay.  But the flux into the mixing region is then entirely due to
diffusion, so this is not a very practical situation.  We will discuss
this case further in Section~\ref{sec:variance}.  Another striking
feature of the curve in Fig.~\ref{fig:bp_vdecay} is that some curves
have a local maximum.

The enhancement to the decay rate is related to shear or Taylor
dispersion, in the sense that it arises from the tilting of
concentration contours.  However, here it is not due to shear, but to
the cross-velocity~$\vc$.

\subsection{Computing the variance}
\label{sec:variance}

For a single Fourier mode, the variance integrated over a
cross-section is
\begin{equation}
  \overline{\theta^2}
  = \int_0^{\Ly}\theta^2\dint\yc = 2\Ls\l\lvert\F(\Xc)\r\rvert^2
  = 2\Ls\l\lvert\F(0)\r\rvert^2\ee^{-2\vdecay\Xc},
  \label{eq:Xsecvar}
\end{equation}
and the total variance in the domain~$\MR = [0,\infty)\times[0,\Ly]$ is
\begin{equation}
  \lavg\theta^2\ravg
  = \int_0^\infty 2\Ls\l\lvert\F(0)\r\rvert^2\ee^{-2\vdecay\Xc} \dint\xc
  = \Ls^2\,\frac{1}{\vdecay}\l\lvert\F(0)\r\rvert^2,
\end{equation}
or in term of the standard deviation
\begin{equation}
  \frac{1}{\Ls \Hflux}\,\lavg\theta^2\ravg^{1/2} = \frac{\Pe}
  {2\sqrt{\vdecay}\sqrt{(\Pe\,\cos\pang + \vdecay)^2 + \omega^2}}\,.
  \label{eq:boxvariance}
\end{equation}
\begin{figure}
  \includegraphics[width=.6\textwidth]{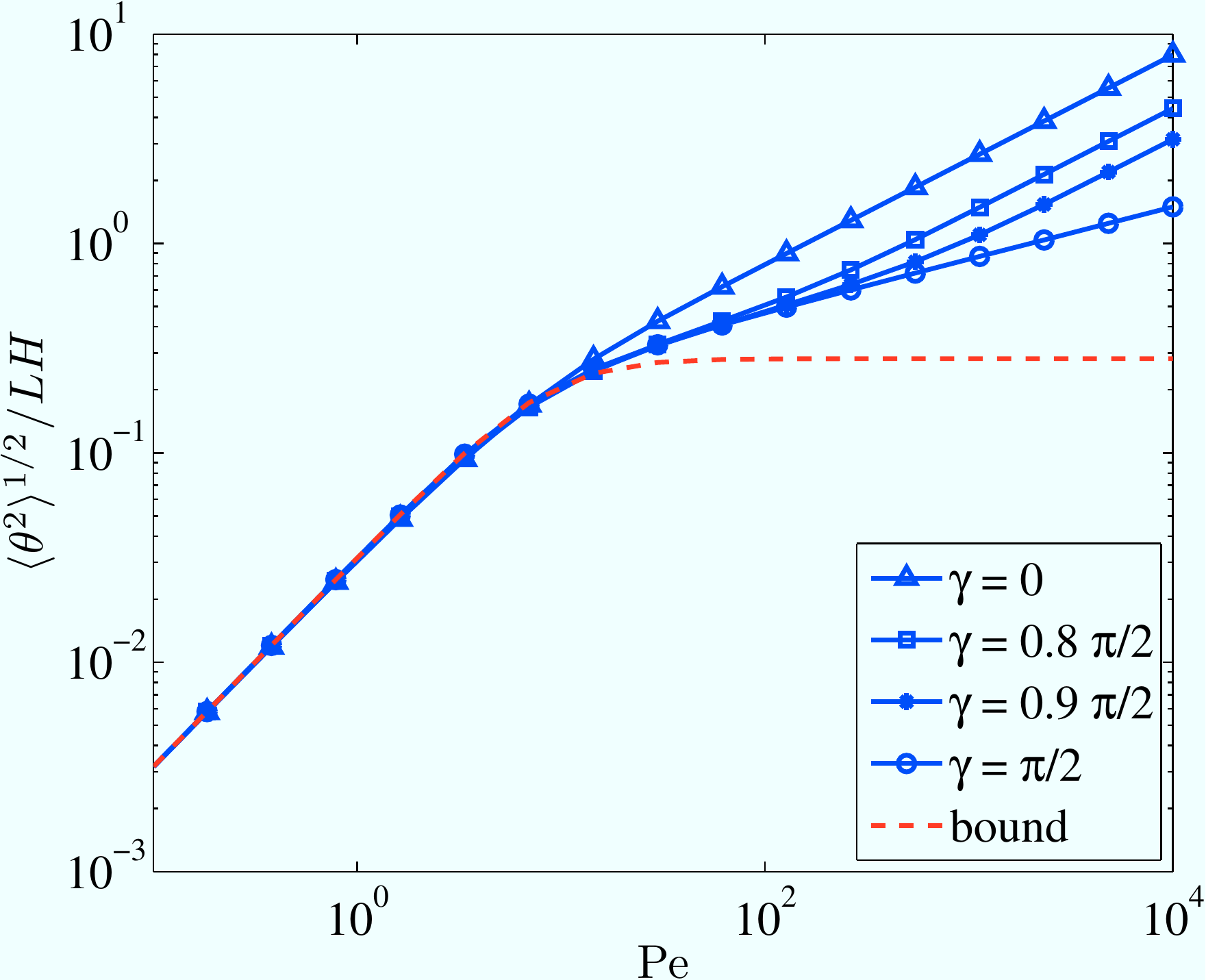}
  \caption{Comparison of the standard deviation of the concentration
    field for the barberpole mixer (solid lines) and the global
    (lower) bound~\eqref{eq:globalboundapprox} for all flows. For
    small~$\Pe$, all the curves are linear in~$\Pe$.  For large~$\Pe$,
    the barberpole solutions scale as~$\Pe^{1/2}$,
    unless~$\pang=\pi/2$ for which the scaling is~$\Pe^{1/4}$.}
  \label{fig:bp_bound}
\end{figure}
For small P\'eclet number, the standard deviation is
\begin{equation}
  \frac{1}{\Ls \Hflux}\,\lavg\theta^2\ravg^{1/2} = \frac{\Pe}{2\Qc^{3/2}}
  - \frac{\Pe^2\,\cos\pang}{8\qc^{5/2}} + \Order{\Pe^3},
  \qquad \Pe\ll 1,
\end{equation}
whilst for large~$\Pe$ it has the form
\begin{equation}
  \frac{1}{\Ls \Hflux}\,\lavg\theta^2\ravg^{1/2} =
  \frac{1}{2\Qc}\,\sqrt{\Pe\,\cos\pang} + \Order{\Pe^{-3/2}},
  \qquad \Pe\gg 1.
  \label{eq:largePe}
\end{equation}
If~$\pang=\pi/2$, then the $\Order{\Pe^{3/2}}$ term in the large $\Pe$
expression diverges, and we have
\begin{equation}
  \frac{1}{\Ls \Hflux}\,\lavg\theta^2\ravg^{1/2} =
  \l(\frac{\Pe}{8\Qc^3}\r)^{1/4} + \Order{\Pe^{-3/4}},
  \qquad \Pe\gg 1,\ \pang=\pi/2.
\end{equation}
These scalings are evident in Fig.~\ref{fig:bp_bound}, which compares
the exact standard deviation~\eqref{eq:boxvariance} to the
bound~\eqref{eq:globalboundnum}.  For small $\Pe$ the exact standard
deviation and the bound are both linear in $\Pe$.  For large $\Pe$ the
lower bound flattens out and becomes independent of $\Pe$.  The
closest the exact standard deviation for the barberpole flow comes to
the theoretical lower bound is for $\pang=\pi/2$, since then the exact
result scales like $\Pe^{1/4}$.  For any other value of $\pang<\pi/2$
the $\Pe^{1/4}$ scaling holds for a short while, but eventually rises
as $\Pe^{1/2}$ when $\Pe\cos\pang$ becomes large.  Hence, the
`optimal' barberpole flow is one for which the horizontal flow~$\Uc$
vanishes ($\pang=\pi/2$): in that case all of the velocity field is
devoted to pushing the source onto the sink, in a manner similar to
the optimal flows found for closed systems~\cite{Shaw2007,
  Thiffeault2008}.  However, unlike the closed systems this flow does
not scale optimally: the bound suggests that there could exist a flow
producing a shallower scaling.  We will discuss the possibility that such
flows exist in Section~\ref{sec:meander}.  Note also that as mentioned
in Section~\ref{sec:barberpoleAD} the flow with~$\pang=\pi/2$ is very
singular: the scalar flux into the domain is entirely due to
diffusion, so there is no \emph{mass} flux into the mixing region.
This is not a very realistic concept on which to build an industrial
mixer.

\section{Example: A Meandering Flow Mixer}
\label{sec:meander}

The barberpole flow of Section~\ref{sec:barberpole} is a nice model
because of its exact solvability but it has some undesirable
pathologies.  Most importantly it is hard to relate it to practical
situations where a channel flow enters and then exits a well-defined
`mixing region,' where mechanical stirring occurs and most of the
mixing takes place.  A better example would involve a flow that varies
in~$\xc$.

As a more general version of the constant barberpole
flow~\eqref{eq:bpflow}, we now consider the case for which the~$\yuv$
component is a functions of~$\xc$,
\begin{equation}
  \uv(\xc) = \uc\,\xuv + \vc(\xc)\,\yuv
  \label{eq:meander}
\end{equation}
with~$\uc>0$.  The domain and boundary conditions are as in
Section~\ref{sec:barberpole}.  This is a `meandering' velocity field
that changes its direction with~$\xc$ but never backtracks.  Once
again we Fourier expand as in~\eqref{eq:Fourierexp} and focus on a
single mode~$\fq_\qc(\xc)$.  The equation to solve is
still~\eqref{eq:ADsinglemode} but with~$\vc(\xc)$ nonconstant.  Unlike
the constant flow case, we cannot solve the equation in full
generality, but since we are mostly interested in the small~$\Diff$
case the WKBJ method allows us to get an explicit approximate
solution.  We assume the standard WKBJ ansatz
\begin{equation}
  \fq_\qc(\xc) = \fq_\qc(0)\,\exp(\Sw(\xc))
  = \fq_\qc(0)\,\exp(\Sw_0(\xc) + \Diff\,\Sw_1(\xc) + \cdots)
\end{equation}
where~$\Sw(0)=0$, and to save notation we left out a~$\qc$ subscript
on~$\Sw$.  After inserting into~\eqref{eq:ADsinglemode} and collecting
terms in~$\Diff$, we can easily solve for the first two terms,
\begin{subequations}
\begin{align}
  \Sw_0(\xc) &= -\imi\,\frac{\qc}{\uc} \int_0^\xc\vc(\ixc)\dint\ixc\,,\\
  \Sw_1(\xc) &= -\frac{\qc^2\xc}{\uc}
  - \imi\frac{\qc}{\uc^2}(\vc(\xc) - \vc(0))
  -\frac{\qc^2}{\uc^3}\int_0^\xc\vc^2(\ixc)\dint\ixc\,.
\end{align}%
\end{subequations}
We will not be finding higher-order terms, and thus
decree~$\Sw=\Sw_0+\Sw_1$; it will prove more convenient to decompose
this into real and imaginary parts as~$\Sw(\xc) = \Swr(\xc) +
\imi\,\Swi(\xc)$, with
\begin{subequations}
\begin{align}
  \Swr(\xc) &= -\frac{\Diff\qc^2}{\uc^3}\l(\uc^2\xc
  + \int_0^\xc\vc^2(\ixc)\dint\ixc\r),\\
  \Swi(\xc) &= -\frac{\qc}{\uc}\int_0^\xc\vc(\ixc)\dint\ixc
  - \frac{\Diff\qc}{\uc^2}(\vc(\xc) - \vc(0))\,.
\end{align}
\end{subequations}
Now we need to apply boundary conditions at~$\xc=0$, where the flux is
\begin{equation}
  \Fx(0,\yc) = \fq_\qc(0)\,(\uc - \Diff\Sw'(0))\,\ee^{\imi\qc\yc} + \cc
  = \Us \Hflux \sin \qc\yc
\end{equation}
where we used~$\Sw(0)=0$.  We use this to solve for
\begin{equation}
  \fq_\qc(0) = \frac{\Us \Hflux}{2\imi\,(\uc - \Diff\Sw'(0))}\,.
\end{equation}
Note that the derivative in~$\Sw'$ simply annuls the integrals in the
definition of~$\Sw$, so this prefactor only involves the unintegrated
velocity field at~$\xc=0$.  The variance integrated over a
cross-section for a single Fourier mode is
\begin{equation}
  \overline{\theta^2}
  = \int_0^{\Ly}\theta^2\dint\yc = 2\Ly\l\lvert\fq_\qc(0)\r\rvert^2
  \ee^{2\Swr(\xc)},
\end{equation}
with
\begin{equation}
  \lvert\fq_\qc(0)\rvert =
  \frac{\Us \Hflux}{2\,\sqrt{(\uc - \Diff\Swr'(0))^2 + (\Diff\Swi'(0))^2}}\,.
  \label{eq:thetaqmag}
\end{equation}
Thus~$\Swr(\xc)$ characterizes the spatial decay of the concentration
field with~$\xc$.  The total, spatially-integrated variance is
\begin{equation}
  \lavg{\theta^2}\ravg
  = 2\Ly\l\lvert\fq_\qc(0)\r\rvert^2\int_0^\infty\ee^{2\Swr(\xc)}\dint\xc.
  \label{eq:var}
\end{equation}

\begin{figure}
  \includegraphics[width=.6\textwidth]{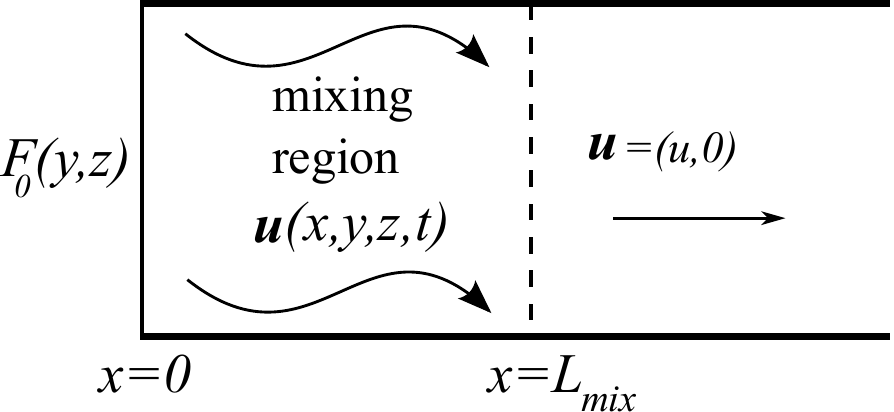}
  \caption{The domain for the meandering flow: within the mixing
    region the flow takes the form~\eqref{eq:meander}, and
    for~$\xc>\Lmix$ it has the simpler, constant
    form~$(\ucinf,0)$.}
  \label{fig:mixingregion}
\end{figure}
To make further progress we will now assume that we have a mixing
region localized in~$\xc$ of length~$\Lmix$, extending from~$0<\xc\le
\Lmix$ (see Fig.~\eqref{fig:mixingregion}).  Before and after that
region we have~$\vc=0$ as well as~$\vc'(0)=0$.  In that
case~$\Sw'(0)=\Swr'(0)=-\Diff\qc^2/\ucinf$.  As for the integral
in~\eqref{eq:var}, it splits as
\begin{equation}
  \int_0^\infty\ee^{2\Swr(\xc)}\dint\xc
  = \int_0^{\Lmix}\ee^{2\Swr(\xc)}\dint\xc
  + \int_{\Lmix}^\infty\ee^{-2\Diff\qc^2\xc/\ucinf}\dint\xc\,.
  \label{eq:intsplit}
\end{equation}
Since~$\Swr\sim\Diff$ is assumed small, the first integral is roughly
equal to~$\Lmix$ and the second integral can be done exactly, yielding
\begin{equation}
  \int_0^\infty\ee^{2\Swr(\xc)}\dint\xc
  \simeq \Lmix
  + \frac{\ucinf}{2\Diff\qc^2}\,\ee^{-2\Diff\qc^2\Lmix/\ucinf}\,.
\end{equation}
We already assumed that~$\Diff\qc^2\Lmix/\ucinf$ was small in
approximating the first integral by~$\Lmix$, so we do it again in the
exponential term to find after expanding
\begin{equation}
  \int_0^\infty\ee^{2\Swr(\xc)}\dint\xc
  \simeq \frac{\ucinf}{2\Diff\qc^2}\,.
  \label{eq:intapprox}
\end{equation}
Neglecting the~$\Diff$ in the denominator of~\eqref{eq:thetaqmag}, we
combine it with~\eqref{eq:intapprox} in~\eqref{eq:var} to find
\begin{equation}
  \lavg{\theta^2}\ravg
  \simeq 2\Ly \frac{\Us^2 \Hflux^2}{4\,\ucinf^2}\,\frac{\ucinf}{2\Diff\qc^2}
  = \Ly \frac{\Us^2 \Hflux^2}{4\,\ucinf\Diff\qc^2}\,.
  \label{eq:meandervar0}
\end{equation}
Recall the definition of~$\Us$, Eq.~\eqref{eq:Usdef}: outside the
mixing region we have~$\csa{\lvert\uv\rvert^2}/\Ly=\ucinf^2$, but
inside we have~$\csa{\lvert\uv\rvert^2}/\Ly\ge\ucinf^2$ because of the
extra stirring motion.  Hence, we put $\ucinf=\civ^{-2}\,\Us$,
with~$\civ\ge1$, and obtain
\begin{equation}
  \frac{1}{\Ls \Hflux}\,\lavg\theta^2\ravg^{1/2}
  \simeq \frac{\civ}{2\Qc}\,\Pe^{1/2}
  \label{eq:meandervar}
\end{equation}
where the length scale~$\Ls=\Ly$ and~$\Qc=\qc\Ly$.  For a constant
flow, we have~$\civ=1$ and we recover the asymptotic
form~\eqref{eq:largePe} with~$\pang=0$.  The constant~$\civ\gtrsim1$
measures the vigor of stirring within the mixing region as compared
with the mean flow.

The standard deviation paradoxically appears to increase with~$\civ$:
this is an artifact of our dimensionless scaling, which was natural
for the bounding approach of Section~\ref{sec:globalbound} but is less
appropriate here.  If instead we normalize the flux as~$\Fx_0(\yc) =
\ucinf \Hfluxinf \sin \qc\yc$, where~$\Hfluxinf$ now measures the flux
normalized by the inflow/outflow velocity~$\ucinf$, we find
\begin{equation}
  \lavg{\theta^2}\ravg
  \simeq 2\Ly \frac{\ucinf^2 \Hfluxinf^2}{4\,\ucinf^2}\,
  \frac{\ucinf}{2\Diff\qc^2}
  = \Ly \frac{\ucinf \Hfluxinf^2}{4\Diff\qc^2}\,,
\end{equation}
or
\begin{equation}
  \frac{1}{\Ls\Hfluxinf}\,\lavg{\theta^2}\ravg^{1/2}
  \simeq \frac1{2\Qc}\,\Peinf^{1/2},\qquad
  \Peinf \ldef \frac{\ucinf\,\Ls}{\Diff}.
  \label{eq:meandervar2}
\end{equation}
So if we use a P\'eclet number based on the inflow, the variance is
completely independent of the stirring flow, at leading order
in~$\Peinf$.  The stirring is too weak to change not just
the~$\Peinf^{1/2}$ scaling, but also the prefactor.  This might seem
surprising but is actually rather obvious: we assumed in calculating
Eqs.~\eqref{eq:meandervar} and~\eqref{eq:meandervar2} that the small
diffusion dominated even in the mixing region.  This is justified
since the flow is 2D and steady, so chaotic advection cannot
occur~\cite{Aref1984}.  Hence the finite mixing region does not
decrease the variance enough, and the total variance is dominated by
the infinite region~$\xc>\Lmix$.

The scaling~\eqref{eq:meandervar} for the general `meandering flow'
shows that there is no hope of approaching the lower bound~$\Pe^0$
scaling with such a flow.  Let us assume that the fluid gets
well-mixed in the mixing region, in a diffusion-independent way as
occurs for chaotic and turbulent mixing.  Then the first term in
integral~\eqref{eq:intsplit} yields a diffusion-independent
constant~$\Lmix$, and the second term can be neglected.  The variance
Eq.~\eqref{eq:meandervar0} becomes
\begin{equation}
  \lavg{\theta^2}\ravg
  \simeq 2\Ly \frac{\ucinf^2 \Hfluxinf^2}{4\,\ucinf^2}\,\Lmix\,,
\end{equation}
which leads to the standard deviation
\begin{equation}
  \frac{1}{\Ls \Hfluxinf}\,\lavg\theta^2\ravg^{1/2}
  \simeq \tfrac12(\Lmix/\Ls)^{1/2}.
  \label{eq:meandervarturb}
\end{equation}
This indeed exhibits a~$\Pe$-independent scaling.  We emphasize that
the difference between~\eqref{eq:meandervarturb}
and~\eqref{eq:meandervar2} is that the former assumes that the length
of mixing region required to reduce the variance to a certain level
becomes independent of the diffusivity as~$\Diff$ becomes smaller,
which is a standard assumption of turbulent and chaotic mixing.
(In fact, typically there will be a weak logarithmic correction
to~$\Lmix$, having to do with the length needed to create small
scales.)  In any case it remains an open challenge to produce an
explicit flow that realizes such optimal mixing.

\section{Summary}

We have proposed a framework for the theoretical investigation of
questions of mixing efficiency in open flows with passive scalar
tracer concentration variations sustained by sources and sinks at an
inlet.  The measure we used to gauge the effectiveness of the stirring
is the spatio-temporal variance of the scalar concentration, a
quantity that can be analyzed via the advection-diffusion equation.
We derived lower bounds on the variance that are observed to be sharp
at low P\'eclet number (weak stirring) and to capture the qualitative
trend at high P\'eclet number (strong stirring) for a particular model
problem, the ``barberpole'' flow.  Concentration variations introduced
at an inlet and subsequently swept downstream are likely most
effectively stirred (for the mixing criterion adopted here) by flows
that, in a downstream moving frame, provide the kind of chaotic mixing
known to work well for transient mixing problems.  It remains an open
question to determine if such a flow can be designed or characterized
in detail even for the simple idealized model considered here.

Many other questions remain open as well.  How does the mixing
efficiency of a given flow vary as the inlet source varies?  How does
the mixing efficiency depend on the length scales we choose to focus
on?  For example the gradient variance or some other norm might just
as well be invoked as the gauge to measure mixing efficiency.  This
issue is known to be important for mixing tracer fluctuations injected
via steady bulk sources and sinks, where the dependence of the mixing
measures on the the P\'eclet number depends explicitly on the length
scales in the sources, the sinks, and those chosen for the mixing
measure~\cite{DoeringThiffeault2006,Shaw2007,Okabe2008}.  These
questions remain for future investigations.

\section*{Acknowledgments}

The authors are grateful to W.~R. Young for helpful discussions, and
to the hospitality of the Institute for Mathematics and its
Applications (supported by NSF).  CRD was supported by NSF under
grants PHY-0555324 and PHY-0855335, and J-LT under grant DMS-0806821.


\end{document}